\documentclass{jpp}
\usepackage{graphicx}
\usepackage{epstopdf, epsfig}
\usepackage{subcaption}
\usepackage{lineno}

\shorttitle{Growth of electron holes in radiation belts}
\shortauthor{Osmane et al.}

\title{Subcritical growth of electron phase-space holes in planetary radiation belts}

\author{Adnane Osmane\aff{1}
  \corresp{\email{adnane.osmane@aalto.fi}},
 Drew L. Turner\aff{2}, Lynn B. Wilson III \aff{3} Andrew P. Dimmock \aff{1}, and Tuija I. Pulkkinen \aff{1}}

\affiliation{\aff{1} Aalto University, Department of Electronics and Nanoengineering, Espoo, 02150, Finland
\aff{2}The Aerospace Corporation, El Segundo, CA 90245, USA. \aff{3} NASA Goddard Space Flight Center, Greenbelt, MD 20771, USA}

\begin{document}

\maketitle

\begin{abstract}
The discovery of long-lived electrostatic coherent structures with large-amplitude electric fields ($1 \leq E \leq 500 $ mV/m) by the Van Allen Probes has revealed alternative routes through which planetary radiation belts' acceleration can take place. Following previous reports showing that small phase-space holes, with $q\phi /T^c_e\simeq 10^{-2}-10^{-3}$, could result from electron interaction with large-amplitude whistlers, we demonstrate one possible mechanism through which holes can grow nonlinearly (i.e. $\gamma \propto \sqrt{\phi}$) and subcritically as a result of momentum exchange between hot and cold electron populations. Our results provide an explanation for the common occurrence and fast growth of large-amplitude electron phase-space holes in the Earth's radiation belts.

\end{abstract}

\section{Introduction}  

\noindent Collisionless plasmas have been known for decades to produce nonlinear coherent structures in the form of isolated equilibrium states \citep{BGK} and turbulent fluctuations \citep{Dupree72, Berman85}. Coherent structures can form as a consequence of phase-space density conservation along particle orbits. By obeying a circulation theorem \citep{Bell67}, vortices in phase-space, similar to Bernstein-Green-Kruskal (BGK) modes, can grow even when eigenmode solutions of the dispersion relation, $\epsilon(k,\omega)=0$, are unavailable for momentum and energy exchange mechanisms \citep{Diamond10}. In a pioneering paper on phase-space density structures, \cite{Dupree72} differentiates states of isolated and turbulent coherent structures in terms of the parameter $\tau_T/\tau_{ac}$. The timescale $\tau_T^{-1}\simeq k \sqrt{q\phi/m}$ is a measure of a trapping time in the potential of an electrostatic structure of average amplitude $\phi$ and wavenumber $k$, whereas the timescale $\tau_{ac}^{-1}\simeq k|\frac{\partial \omega}{\partial k}-\frac{\omega}{k}|$ is a measure of a wave-packet dispersal rate as seen by a resonant particle of velocity $v\simeq \omega/k$. In the limit where $\tau_T/\tau_{ac}\ll 1$, resonant particles have enough time to close a trapped orbit and an isolated BGK equilibrium can form. In the opposite limit where $\tau_T/\tau_{ac}\gg 1$, the field pattern changes prior to a particle bouncing in the field of the structure, and the coherent BGK vortices become sheared apart before formation. In such a regime, the plasma state can be characterized by small-scale phase-space structures, termed granulations by \cite{Dupree72}. A turbulent state can then arise when granulations are sufficiently large and collisions between individual structures possible. 

\noindent Following \cite{Dupree72, Dupree82, Dupree83}, and of particular interest to our analysis, \cite{Berman85} showed numerically that highly intermittent fluctuations consisting of small-scale phase-space holes could be produced from an homogeneous ion-electron plasma stable to current-driven ion acoustic instabilities. The subcritical growth of isolated coherent structures could be accounted for by momentum exchange between the ion holes and reflecting electrons \footnote{See more recent numerical studies by \cite{Lesur13} for a discussion on the limits of \cite{Berman85} results.}. In order to conserve phase-space density, the holes grew in amplitude by moving to regions of higher average phase-space density; a mechanism we will argue can also take place in the Earth's radiation belts for electron holes. Later studies clarified the ideas proposed by \cite{Dupree72} and extended the description of self-trapped phase-space structures to magnetized plasmas \citep{Dupree75, Infield77,Misguich78, Boutros81, Dupree82,Berman82, Dupree83, Berman85, Barnes85, Tetreault88, Biglari, Hamza93, Lesur13, Hutchinson15}\footnote{For an introductory account of phase-space turbulence, the reader is referred to \cite{Diamond15} and to chapter 8 of \cite{Diamond10}. Important caveats on the role of small-scale phase-space structures in the context of large-scale plasma turbulence can be found in sections 4.4 and 10.5 of \cite{Krommes} with references therein.}.

\noindent Observations in weakly collisional space plasmas of nonlinear holes associated with large field magnitudes, and their impact on charged-particle acceleration, are not new \citep{Temerin82, Bale98, Ergun98, Franz,Cat02, Ergun02, Bale02, Wilson07, Kellog10, Ergun16, Jaynes16}. For the well-documented case of phase-space holes observed in the Earth's auroral zone, it was inferred from momentum balance between ions and electrons that aggregates of (electron) holes could grow and lead to the heating of the ion population \citep{Ergun98, Ergun02}. But measurements by the Van Allen probes of double layers and electron solitary hole trains, with electric field amplitudes ranging between ($1 \leq E \leq 500 $ mV/m), and properties akin to electron acoustic modes \citep{Mozer13, Malaspina14}, hinted at novel routes for \textit{electron}, not ion, transport and acceleration in planetary magnetospheres. Nonlinear coherent structures were postulated by \cite{Mozer13} as key ingredients in the injection of several keV electrons, with the broader implication that phase-space holes could provide the means through which collisionless astrophysical plasmas can self-consistently support parallel electric fields and acceleration of both ion and electron populations.   

\noindent While the global impact of electron holes in the generation of suprathermal electrons in the radiation belts remains to be quantified theoretically and observationally \citep{Mozer16, Ma16}, an understanding of the wave-particle and nonlinear interactions of coherent electrostatic structures is starting to form. \cite{Osmane14b} and \cite{Art14c} have shown that electron holes can efficiently energize thermal electrons through Landau resonance to keV energies on time scales of a few bounce periods $\tau_b \sim 1-10$s in the Earth's magnetic field. \cite{Vasko16} explained the evolution and associated adiabatic heating of electron holes in an inhomogeneous magnetic field. \cite{Agapitov15} provided observational evidence for the generation of nonlinear coherent structures following the interaction of two large-amplitude whistlers. Showing similar interplay between whistlers and nonlinear coherent structures in the radiation belts, \cite{Osmane16} postulated that Landau-resonant electrons, trapped or scattered into the loss-cone by large-amplitude waveforms, could result in small-scale phase-space structures in the shape of double layers and electron solitary holes. However, this mechanism could only result in small amplitude fluctuations $q\phi/T_e^c \leq 0.01$, since the energetic electron density, $n_h$, only accounted for a small portion of the total electron density, $n_h/(n_c+n_h) \simeq 0.1-1\%$ \citep{Spjeldvik77, Denton10}, and precipitated electrons only accounted for a fraction of the hot and cold electron density. Based on observations \citep{Mozer13, Malaspina14}, large electron holes $q\phi/T_e^c \simeq 1$, with potential energy of the order of the cold electron temperature $T_e^c$, can form in the radiation belts. Consequently growth of phase-space structures from an initial seed fluctuation of the order of the hot to cold electron density $n_h/n_c \simeq 10^{-3}$ is required. Following the argument by \cite{Dupree72} summarized above, we would like to understand how nonlinear structures with electron acoustic properties in the radiation belts evolve from a regime in which $\tau_T/\tau_{ac}\gg 1$ (small to moderate amplitude, i.e. $q\phi/T_e^c \le 0.01-0.001$) to one where $\tau_T/\tau_{ac}\simeq 1$ (large amplitude structures, i.e., $q\phi/T_e^c \geq 1$)\footnote{The auto-correlation times $\tau_{ac}$ is computed from the electron acoustic dispersion relation derived in \cite{Gary85, Gary87}.}.

\noindent In the following report, we derive a heuristic expression for the nonlinear growth rate of electron holes for parameters consistent with the Earth's radiation belts. In sections 2.1 we discuss the seed fluctuation scaling of electron phase-space structures. In sections 2.2 we summarize the  formalism used to compute the growth rate of phase-space structures.  In 2.3 we describe the mean field theory used to quantify the interaction of hot electrons with electrostatic localized structures. In 2.4 we compute growth rates for a parameter space consistent with radiation belt plasmas and compare our results with the linear theory of the electron acoustic mode in similar conditions. While our analysis is derived for the simplified case of unmagnetized, electrostatic and homogeneous plasma, it provides a simplified illustration of momentum-exchange mechanisms responsible for the growth of electron holes. The same mechanism can be generalized to magnetized and inhomogeneous plasmas.

\section{Dynamics of phase-space electron holes}

\subsection{Seed formation of electron holes in the radiation belts} 
\noindent In a previous report by \cite{Osmane16}, it was argued that seed fluctuations for double layers and electron solitary holes generated after energetic electrons' ($E\geq 1 keV$) interaction with large-amplitude whistler waves scaled as $q\phi/T_e \simeq 10^{-2}-10^{-3}$. We here quantify the scaling of the seed fluctuations in the distribution function resulting from such interaction with large-amplitude whistlers. The formation of phase-space depletions $\delta f<0$ in the distribution function, with $|\delta f|/f_0 \ll 1$, could either be the result of electron scattering in the loss-cone near mirror points, or/and particle trapping in the potential of large-amplitude whistlers. The size of the hole can be estimated by considering conservation of particle orbits for Landau resonant and non-resonant electrons as follows. The Hamiltonian  describing the wave-particle interaction with a localized electrostatic structure in an inhomogeneous magnetic field can be written as : 
\begin{equation}
H=\frac{m_e v_\parallel^2}{2}+\mu B(\mathbf{x})-q\phi_w(\mathbf{x},t),
\end{equation}
where $\mu=m_e v_\perp^2/2B$ is the first adiabatic invariant and $\phi_w$ is the electrostatic potential of the large-amplitude whistler\footnote{The conservation of the first adiabatic invariant follows from ignoring cyclotron-resonant electrons.}. The energy change for the charged particle can similarly be written as: $\frac{dH}{dt}=-\frac{q\partial \phi_w}{\partial t}$. We now envision a Landau resonant electron interacting with a localized parallel electric field $k_\parallel L \gg 1$, where $L^{-1}\simeq \nabla B/B$ is the curvature of the mean field and $k_\parallel$ is the inverse of the length upon which the electron interact strongly with the localized fields \footnote{Note that even if the electrostatic structure has a wavelength $\lambda \simeq L$, the inhomogeneous mean field will constrain the interaction time.}. In the absence of a localized electric field, the particle bounces back and forth between symmetrical mirror points: $v_\parallel^2=\frac{2\mu B_2}{m_e}\left(1-B_1/B_2\right)$, where the subscripts 1 and 2 indicate initial and final position. In the presence of an electric field, the bounce motion between the mirror points will loose its symmetry, and conservation of energy entails the following relation :   $v_\parallel^2=\frac{2\mu B_2}{m_e}\left(1-B_1/B_2\right)-2\Delta E/m_e$, where $\Delta E =-q\int_0^{\tau_b} \frac{\partial \phi_w}{\partial \tau}d\tau$. Gain ($\Delta E>0$) and loss ($\Delta E <0$) of energy can lead to violation of the second adiabatic invariant $J=\oint p_\parallel d s_\parallel$, with Landau-resonant particles either confined to lower latitudes (with $\Delta E<0$ leading to $B_1>B_2$), or pushed to higher latitudes (with $\Delta E>0$ leading to $B_2>B_1$), and in some instances, for sufficiently large electric field amplitudes, into the loss-cone \citep{Osmane16}. In the case where the Landau-resonant hot electrons ($v\simeq v_\phi \simeq 30 000 $ km/s$\gg v_{tc}\simeq 3000$ km/s) experience an energy jump $\Delta E \simeq 0.1-100$ keV, the reduced parallel distribution function of the hot electrons can be computed from the Hamiltonian $H$ and the cross-sectional radius $\bar{v}_\perp=\sqrt{\left(v_\parallel^2-2\Delta E/ m_e\right)/{B_2/B_1-1}}$ by the following integral : 
\begin{eqnarray}
f_h(v_\parallel)&=&\frac{n_{h}}{\sqrt{\pi}v_{th\parallel}v_{th\perp}^2}\int_{\bar{v}_\perp}^\infty dv_\perp \exp\left[-\frac{v_\parallel^2}{v_{th\parallel}^2}-\frac{v_\perp^2}{v_{th\perp}^2}\right] \nonumber \\ 
&=&\frac{n_{h}}{\sqrt{\pi}v_{th\parallel}}\exp\left[-\frac{v_\parallel^2}{v_{th\parallel}^2}-\frac{v_\parallel^2-2\Delta E/m_e}{(B_2/B_1-1)v_{th\perp}^2}\right].
\end{eqnarray}    
Consequently, the shift of the mirror point for Landau resonant particles will lead to either a clump ($\delta f > 0$) or a hole ($\delta f < 0 $) in the reduced distribution function. The perturbation in the distribution function near the mirror point ($B_2\simeq B1$) will scale as $\delta f/f_c=\left[f_h(v_\parallel^2\simeq 0+2\Delta E/m_e)-f_h(v_\parallel^2\simeq 0)\right]R(\Delta v_T)/f_c\simeq n_h\left[1- \exp(-2\Delta E/m_e v_{\parallel t}^2)\right] R(\Delta v_T)/n_c$, where the subscripts $h$ and $c$ stand for hot and cold components and the resonance function $R(\Delta v_T)=\frac{1}{\sqrt{\pi}}\frac{v_{t\parallel}}{\Delta v_T}exp[-(v_\parallel(t)-v_\phi)^2/\Delta v_T^2]$  approximates into a delta function $\delta(v_\parallel(t)-\omega/k_\parallel)$ as the trapping width $\Delta v_T\longrightarrow 0$. The resonance function overlaps with Landau-resonant electrons in the distribution of hot electrons and broadens for larger electric field amplitude. The reduced distribution can be integrated numerically for finite $\Delta v_T$ along the parallel velocity to obtain the number density of particles forming a hole or a clump in the vicinity of a mirror point. For energy gains (losses) much smaller than the temperature of the hot electrons, $\Delta E \ll T_h\simeq m_e v_{th\parallel}^2$, a hole (clump) will scale as:
\begin{equation}
\frac{ \delta f}{f_c}\simeq -\frac{1}{\sqrt{\pi}}\frac{n_h}{n_c}\frac{\Delta v_T}{v_{th\parallel}}<0 (>0).
\end{equation} 
However, for parallel electric fields amplitude $E_\parallel \simeq 10$ mV/m, parallel wavevector $k_\parallel \simeq 2\pi/R_E$, Earth radii $R_E\simeq 6000$ km, wave-frequency $\omega\simeq 3$kHz and interaction time $\Delta t\simeq 1-10$ ms \cite{Art13, Osmane16}, we find that energy gains for Landau resonant electrons scale as $\Delta E/T_h \simeq 10-100$, that is energy gains of the order of 1 to 100 keV for temperature ratios $T_h/T_h\simeq 10$. In this limit $\Delta E \gg T_h$, and the perturbation in the distribution function of the hot electrons will scale as : 
\begin{equation}
\label{scaling1}
\frac{\delta f}{f_c}\simeq -\frac{1}{\sqrt{\pi}}\frac{n_h}{n_c}\frac{v_{th\parallel}}{\Delta v_T}.
\end{equation}
For large-amplitude whistlers, $\Delta v_T/v_{th\parallel} \simeq \sqrt{q\phi_w/T_h}\simeq 0.1-10$, and we find that the distribution fluctuation $\delta f$ is essentially controlled by the ratio of hot to cold densities. In the Earth's radiation belts the ratio of hot to cold density is very small, i.e., $n_h/n_c \simeq 10^{-2}-10^{-3}$, and therefore strongly constrains the size of holes generated by Landau-resonant electrons, even in the case of very large-amplitude whistlers \citep{Art13}. Using the self-consistent Poisson equation in 1 dimension one can write : $-\frac{\partial^2 \phi}{\partial s^2}+\frac{\phi}{\lambda^2}=4\pi q\int_{-\Delta v_s}^{\Delta v_s}\delta f dv\simeq -4\pi q \delta f \Delta v_s$. For shielding distances $\lambda \geq \lambda_D$ in the quasi-neutral approximation ($\partial^2 \phi/\partial s_\parallel^2\ll \phi/\lambda^2$) and phase-space holes with velocity width $\Delta v_s^2 \simeq 2q \phi/m_e$, not to be confused with $\Delta v_T$ which correspond with the width of particles accelerated by the parallel whistlers. We therefore estimate that electrostatic potential amplitudes arising in the well in of large-amplitude whistlers would scale as:
\begin{equation}
\label{scaling2}
 \frac{|q\phi|}{T_c}\simeq\frac{\lambda^2}{\lambda_D^2}\frac{|\delta f|}{f_0}\frac{\Delta v_s}{v_{tc}}\simeq \frac{\lambda^2}{\lambda_D^2}\frac{n_h}{n_c}\left(\frac{T_h}{T_c}\right)^{\frac{1}{2}}\left(\frac{\phi}{\phi_w}\right)^{\frac{1}{2}}\simeq 0.01.
\end{equation}
 
\noindent With the density of Landau-resonant electrons known to be modest when compared to the bulk plasma density, the seed creation of phase-space holes can only lead to small amplitude fluctuations in the electric potential. Since nonlinear electrostatic structures in the radiation belts are also observed with much larger amplitudes $q\phi/T_e\ge0.1-1$ and $\delta f/f \simeq 30$\%, the source of free energy for the growth in the Earth's radiation belts remains unclear. In the following, we will use the above scaling to argue that the growth of phase-space holes could take place as a result of momentum exchange between hot electrons and the phase-space hole. As the hot electron population gains momentum through the interaction with double layers and electron solitary holes, the phase-space hole loses momentum but increases its amplitude in order to conserve phase-space density \citep{Dupree83}. The mechanism operates in similar fashion to current-driven ion instabilities in the linear regime and phase-space hole growth formation for ion-electron plasma \citep{Diamond10}.

\subsection{Phase-space structures' growth and momentum exchange} 
\noindent For completeness, we here provide a derivation of the equation relating momentum exchange between particle species and phase-space structure growth. We note, however, that similar derivation can already be found in the literature \cite{Dupree82, Diamond10, Lesur13}. We start with the Vlasov equation in one dimension for a particle specie $s$: 
\begin{equation}
\frac{\partial f_s}{\partial t}+v\frac{\partial f_s}{\partial x}+\frac{q_s}{m_s}E\frac{\partial f_s}{\partial v}=0.
\end{equation} 
The evolution of the electric field can be computed from Maxwell's equation in the absence of magnetic field as : 
\begin{equation}
\frac{\partial E}{\partial t}=-4\pi\sum_s q_s\int_{-\infty}^{+\infty} v f_s dv  -\gamma_d E.
\end{equation}
The last term on the right-hand side describes the damping of the electric field due to the quasi-linear interaction of the electrostatic turbulence with the resonant electrons \citep{Kadomtsev71}. We now split the distribution function in terms of an ensemble-averaged quantity $f_{os}=\langle f_s\rangle$ and a phase-space disturbance $\delta f_s$ with zero mean, i.e. the plasma is permeated of both clumps $\delta f> 0$ and holes $\delta f <0$. We note that while the fluctuations in the phase-space distribution are not eigenmodes of the linearized systems, they ought to be consistent with Poisson's equation. Since phase-space density must be conserved, the perturbation to the background distribution can be written as : 
\begin{equation}
\frac{d}{dt}\delta f_s=-\frac{q_s}{m_s}E\frac{\partial f_{os}}{\partial v}
\end{equation}
We now multiply the above equation by $\delta f_s$ and integrate over velocity and space to find the following evolution equation for the quantity $\psi_s=\int_{-\infty}^{+\infty}dv\langle \delta f_s^2\rangle$: 
\begin{equation}
\frac{d}{dt}\psi_s=-2\frac{q_s}{m_s}\int_{-\infty}^{+\infty}dv\langle E \delta f_s\rangle \frac{\partial f_{os}}{\partial v}
\end{equation} 
For phase-space structures with small velocity width, i.e., $\delta f_s \in  [u_s-\Delta v_T, u_s+\Delta v_T]$ such that we can approximate $\frac{\partial f_{os}}{\partial v}\simeq \frac{\partial f_{os}}{\partial v}\big{|}_{u_s}$ over the interval  $[u_s-\Delta v_T, u_s+\Delta v_T]$, the above integral takes the form : 
\begin{equation}
\frac{d}{dt}\psi_s\simeq-\frac{2}{m_s}\frac{\partial f_{os}}{\partial v}\bigg{|}_{u_s}\int_{-\infty}^{+\infty}dv\langle q_sE \delta f_s\rangle. 
\end{equation}
The integral on the right-hand side is the rate of change of momentum for particle specie $s$, hence, the evolution of the phase-space structure $\psi_s$ is linked to the momentum of specie $s$ through the following integral : 
\begin{equation}
\label{master_eq}
\frac{d}{dt}\psi_s\simeq-\frac{2}{m_s}\frac{\partial f_{os}}{\partial v}\bigg{|}_{u_s} \frac{d}{dt}\langle p_s\rangle.
\end{equation}
\noindent To the best of the authors' knowledge, this equation was first derived by \citep[See equation 12, p280]{Dupree82}. As described by \citep{Diamond10}, this equation is remarkable because it indicates how a local structure $\langle\delta f_s^2\rangle$ in the distribution function is linked to the evolution of net momentum transfer. For such growth to take place, no linear modes are necessary, and momentum exchange can take place between two species or components.
For two particle species $s=i,e$, and a phase-space hole in the ion distribution, i.e., $\delta f_i \neq 0$, a phase-space hole can grow (damp), if the electron distribution can gain (loose) momentum $dp_e/dt=-d \langle p_i\rangle/dt$ $>0$ ($<0$). The growth (damping) of a phase-space structure is a consequence of conservation phase-space density along particle orbits. As the phase-space hole gives (gains) momentum away, it recedes to regions of higher (lower) phase-space density and grows (damps) to preserve phase-space volume. One can complement the above equation for the phase-space structure with an evolution equation for the electrostatic wave energy density $W=\langle E^2 \rangle/4\pi$ :
\begin{equation}
\frac{d}{dt}W=-2\gamma_d W-\sum_sq_s\int_{-\infty}^{+\infty}v\langle E\delta f_s\rangle dv.
\end{equation} 
The second term on the right-hand side is always neglected in quasi-linear theory since it is of order $\delta f_s ^2$. This second term takes into account the bunching effect due to particle trapping and can lead to wave amplification despite linear wave dissipation $\gamma_d>0$ \citep{Kadomtsev71}. 

\subsection{Mean-field theory of the hot electron population} \noindent We model the impact of localized electrostatic structures on the hot electron population by using the mean-field formalism initially derived by \cite{Morales} for an electrostatic turbulence composed of solitons and extended by \cite{Melrose89} for the more general case of inhomogeneous distribution of electrostatic structures. The scattering experienced by electrons results in a Fokker-Planck equation dependent on average localized electric fields of amplitude $E$ and scale $d \sim \lambda_D$. The quasi-linear approach to the problem applies provided that an electron has multiple encounters with localized structures and that the relaxation of the background distribution function occurs on timescale much longer than the average interaction time. In the derivation of \cite{Morales} suprathermal electrons do not participate in the formation of the localized electrostatic structures, nor interact strongly with them. We note that this assumption is also valid for the hot electrons in the radiation belts and consistent with the analytical and numerical estimates of \cite{Osmane14b}. Consequently, the temporal evolution of the hot electron distribution, $f_{0h}=f_{0h}(v,t)$, in our problem is determined by the following diffusion equation: 
\begin{equation}
\label{Morales_eq}
\frac{\partial f_{0h}}{\partial t}=\frac{\partial}{\partial v}\bigg{(}D(v)\frac{\partial f_{0h}}{\partial v}\bigg{)},
\end{equation}
with the velocity dependent coefficient scaling as:
\begin{equation}
D(v)=\frac{\pi}{\bar{\tau}}\frac{d^2}{\lambda_{D}^2}\frac{|E^2|}{4\pi n_cT_c}\frac{\omega^2}{\omega_{pe}^2}\frac{v_{th,c}^3}{|v|}\cosh^{-2}\bigg{(}\pi\frac{d}{\lambda_{D}}\frac{v_{tc}}{v}\bigg{)}.
\end{equation}
In the above, the diffusion coefficient is written in terms of the Debye length $\lambda_{D}=\sqrt{T_c/4\pi n_c e^2}$ (from now on we drop the subscript $e$ and all temperatures refer to electrons), the average size of a electrostatic structure $d$, the frequency of the seed fluctuation $\omega$, the plasma frequency $\omega_{pe}=\sqrt{4\pi n_cq^2/m_e}$, the spatially averaged electric field intensity $|E^2|$, and the thermal speed of the cold electron component $v_{tc}=\sqrt{2T_c/m_e}$. The time scale $\bar{\tau}$ refers to a mean free-path for the collision between energetic electrons and electrostatic structures. In the case of a plasma efficiently confined by an inhomogeneous magnetic field, average collision times would scale as the bounce period $\tau_b$ for a single localized field along the field line, half a bounce period $\tau_b/2$ for two localized fields, and similarly for a number $N$ of localized fields, i.e. $\tau_b/N$. We note that the diffusion coefficient is symmetric with respect to the axis $D(0)=0$, peaks at $D\simeq  \frac{|E^2|}{2m_e n_c\bar{\tau}}$ for speeds $|v|\simeq \frac{2}{3}\frac{\pi d}{\lambda_{D}}v_{th}$, is of order $e^{-2d/\lambda_D}\sim10^{-4}$ smaller for $|v|\simeq v_{th}$, and rapidly converges to zero for $v \rightarrow 0$. For $d\geq 4 \lambda_D$, supra-thermal particles with $|v| \sim 10 v_{th}$ interact the most strongly with the localized electric fields. It should be pointed out, that the time-dependent solutions of Equation (\ref{Morales_eq}) also exhibit suprathermal tails for an initially Maxwellian distribution. As detailed further below, our choice of diffusion coefficient constrains momentum exchange with the hot electron population whenever a drift between the hot and cold electrons is present. The inclusion of the cold electron dynamics requires additional care. Cold electrons can be trapped and reflected by the localized electrostatic structures in the radiation belts and a quasi-linear formalism to describe their associated transport would be inappropriate. However, as we will show below, simple consideration of momentum exchange between phase-space holes, situated in the cold electron distribution, and the hot electrons who interact weakly with the fields, can nonetheless lead to structure growth for subcritical regime \footnote{A complete treatment including the cold electrons, trapped and passing, in a magnetized and inhomogeneous plasma, has been left for future studies \citep{Osmane17}. Our basic result, however, remains. Momentum exchange between hot and cold electrons can lead to the subcritical growth of electrostatic structures in planetary radiation belts.}.     

\subsection{Application to the Earth's radiation belt}

\noindent In a plasma with two electrons population, like in the Earth's radiation belts, we can track momentum-exchange associated with cold electrons $p_e^c$, hot electrons $p_e^h$ and ions $p_i$ in terms of the following conservation law : 
\begin{equation}
\frac{d}{dt}\big{(}\langle p_e^c \rangle+p_e^h+p_i\big{)}=0,
\end{equation}
where $\langle p_e^c\rangle$ is the coherent structure averaged momentum, i.e. $\langle p_e^c\rangle/m_e=\int dv (v-u)\delta f_c$, with the structure centered at velocity $u$. For an ion temperature $T_i\simeq 1$ keV \citep{Spjeldvik}, and cold electron temperature $T_c \simeq 30-100$ eV, the ion thermal speed scales as $20-40\%$ of the electron thermal speed $v_{tc}$. Since the electron phase-space structures in the Earth's radiation belts propagate at speeds $v_\phi \ge v_{tc} \gg \sqrt{T_i/m_i}$, ions are far from resonances, and we safely assume that the ion net momentum change can be ignored, i.e., $d p_i/dt \simeq 0$. For the sake of simplicity, the wave (fields plus non-resonant cold electrons) momentum is, in quasi-linear fashion \citep{Diamond10, Kadomtsev71}, assumed to balance the resonant particle kinetic momentum. 

\noindent These simplifications allow us to quantify the momentum exchange between the diffuse hot electrons and an initially small depletion $|\delta f|/f \ll 1$ in the cold electron distribution function : a mechanism not included in quasilinear theory concerned solely with the spatially-averaged distribution functions.

\noindent We therefore write the distribution for the cold electrons in terms of a background component $f_{0c}$ and a localized structure $\delta f_c$, i.e., $f_e^c=f_0^c+\delta f_c$. Making use of Equation (\ref{master_eq}), we can write : 
\begin{equation}
\label{maineq}
\frac{d}{d t}\int_{-\infty}^{+\infty} dv \ \langle \delta f_c ^2\rangle=\frac{2}{m_e}\frac{\partial f_{0c}}{\partial v}\bigg{\arrowvert}_u \frac{d p_e^h}{dt}.
\end{equation}
Using the diffusion equation derived by \cite{Morales} and \cite{Melrose89} and integration by parts, the momentum of the hot electron population can be written as :
\begin{equation}
\frac{d p_e^h}{dt} =-m_e\int_{-\infty}^{+\infty} dv\ D(v)\frac{\partial f_e^h}{\partial v}.
\end{equation}
Equation (\ref{maineq}) therefore takes the following form :
\begin{equation}
\label{maineq}
\frac{d}{dt}\int_{-\infty}^{+\infty} dv\ \langle \delta f_c ^2\rangle=-2\frac{\partial f_{0c}}{\partial v}\bigg{\arrowvert}_u\int_{-\infty}^{+\infty} dv\ D(v)\frac{\partial f_0^h}{\partial v}.
\end{equation}

\noindent This equation can be solved exactly for traditional quasi-linear diffusion coefficients of Langmuir waves \citep{Diamond10}, but requires asymptotic analysis or numerical integration when dealing with the diffusion coefficient of \cite{Morales} and \cite{Melrose89}. We now assume that the cold and hot electron distributions are Maxwellian of the form $f_0^s=n_s\sqrt{m_e/2\pi T_s}\exp(-m_e(v-v_d^s)^2/T_s)$, for cold and hot electron species $s=c,h$, with density $n_s$, thermal velocity $v_{ts}$ and temperature $T_s=m_e v_{ts}^2$ and drift velocity $v_d^s$. If we only set the hot component population with a drift velocity in equation (\ref{maineq}), that is, we compute the growth rate in the frame of the cold electron population, we find the following expression : 
\begin{eqnarray}
\label{maineq2}
\frac{d}{dt }\int_{-\infty}^{+\infty} dv\ \langle \delta  f_c ^2  \rangle &=&-4\frac{n_cn_h}{\bar{\tau}v_{th}}\frac{d^2}{\lambda_{D}^2}\frac{\omega^2}{\omega_{pe}^2}\frac{u}{v_{th}}\frac{|E^2|}{4\pi n_cT_c}\exp\bigg{(}-\frac{u^2}{v_{tc}^2}\bigg{)}\nonumber \\ 
&\times&\int_{-\infty}^{+\infty} \frac{dv}{v_{th}} \big{(}\frac{v}{|v|}-\frac{v_d^h}{|v|}\big{)}\cosh^{-2}\bigg{(}\pi\frac{d}{\lambda_{D}}\frac{v_{tc}}{v}\bigg{)} \exp\bigg{[}\frac{-(v-v_d^h)^2}{v_{th}^2}\bigg{]}.
\end{eqnarray}
We write the normalized electric field density in terms of the electric potential as $|E^2|/4\pi n_c T_c =|q\phi/T_c|^2k^2\lambda_D^2$ and use equations (\ref{scaling1}) and (\ref{scaling2}) to determine the scaling of the electric potential as: 
\begin{equation} 
\frac{q\phi}{T_e} \simeq \left[\frac{\delta f_c}{f_{0c}}\frac{\lambda^2}{\lambda_D^2}\right]^2. 
\end{equation}
By approximating the left-hand side integral as $\gamma \delta f_c^2 \Delta v$, for which we assumed $\langle\delta f_c^2\rangle \simeq \exp(\gamma t)$, dividing both sides by $f_{0c}^2 v_{th}$, and assuming a shielding distance of the order of the Debye length,
we find that the growth rate of the seed fluctuations to be estimated by the following expression: 
\begin{eqnarray}
\label{maineq3}
\gamma &=& -\frac{4}{\bar{\tau}}\left(\frac{n_h}{n_c}\right)\left(\frac{T_c}{T_h}\right)^{\frac{3}{2}}\frac{u}{v_{tc}}\sqrt{\frac{|q\phi|}{T_c}}\exp\bigg{(}-\frac{u^2}{v_{tc}^2}\bigg{)} k^2 d^2 \nonumber \\ 
&\times&\int \frac{dv}{v_{tc}} \left(\frac{v}{|v|}-\frac{v_d^h}{|v|}\right)\cosh^{-2}\bigg{(}\pi\frac{d}{\lambda_{D}}\frac{v_{tc}}{v}\bigg{)} \exp\bigg{[}\frac{-(v-v_d^h)^2}{v_{th,h}^2}\bigg{]}.
\end{eqnarray}
And thus, the growth or damping rate of the structure is nonlinear, since it is proportional to the square root of the fluctuation, i.e., $\gamma \sim  \sqrt{q\phi/T_c}$. Finally, we write the final form of the instability growth rate of the electrostatic structure in a more compact form in terms of the normalized variables : $\xi=v/v_{tc}$,  $\chi=T_c/T_h$, $x=u/v_{tc}$, $\bar{v}_d=v_d^h/v_{tc}$, $\Phi=q\phi/T_c$, $\bar{n}=n_h/n_c$, as
\begin{eqnarray}
\label{maineq4}
\gamma = \alpha\frac{\sqrt{\Phi}}{\bar{\tau}}\int d\xi\bigg{(}\frac{\xi}{|\xi|}-\frac{\bar{v}_d}{|\xi|}\bigg{)}\cosh^{-2}\bigg{(}\pi\frac{\eta}{\xi}\bigg{)} \exp\big{[}-(\xi-\bar{v}_d)^2\chi\big{]}.
\end{eqnarray}
In the above, the parameter $\alpha= -4\bar{n}\chi^{3/2}x\exp(-x^2)$ is a function of the temperature and density ratios between hot and cold electron populations. In the following we set $x\simeq 1/\sqrt{2}$, indicating that the electron hole velocity is of the order of the thermal electron velocity, and that therefore $\alpha<0$. The case for $x<0$, $\alpha>0$, will only be described for the sake of brevity but can easily be generalized by reversing signs of the growth/damping rate $\gamma$ when switching from one case to the other. Solving the integral numerically we can now determine the parameter values for which structures grow or damp.

\subsection{Dependence of structure growth on plasma parameters }
\begin{figure}
\begin{subfigure}{.5\textwidth}
  \centering
  \includegraphics[height=5.2cm,width=4.7cm]{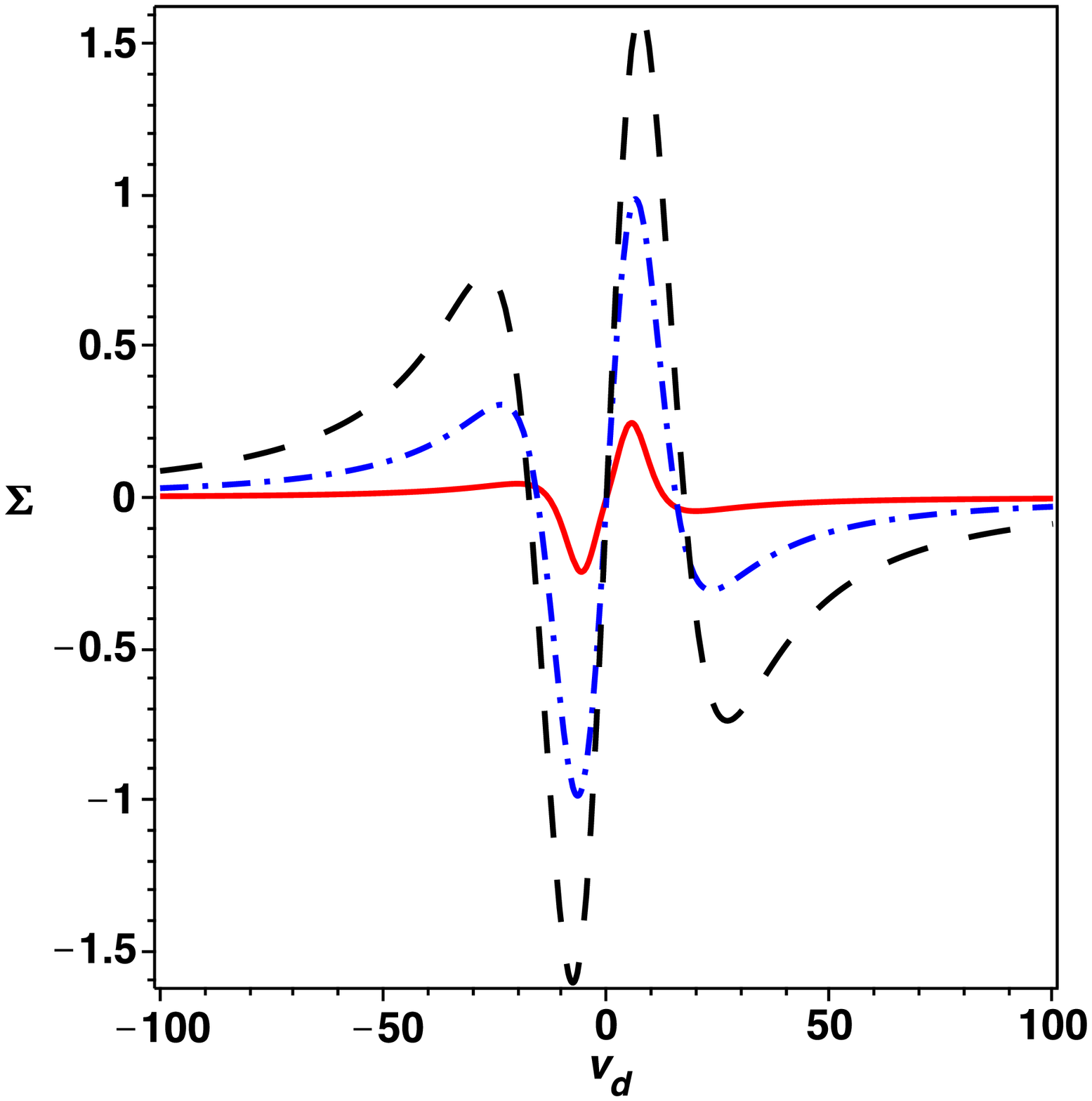}
  \label{fig:sfig1}
\end{subfigure}%
\begin{subfigure}{.5\textwidth}
  \centering
 \includegraphics[height=5.2cm,width=4.7cm]{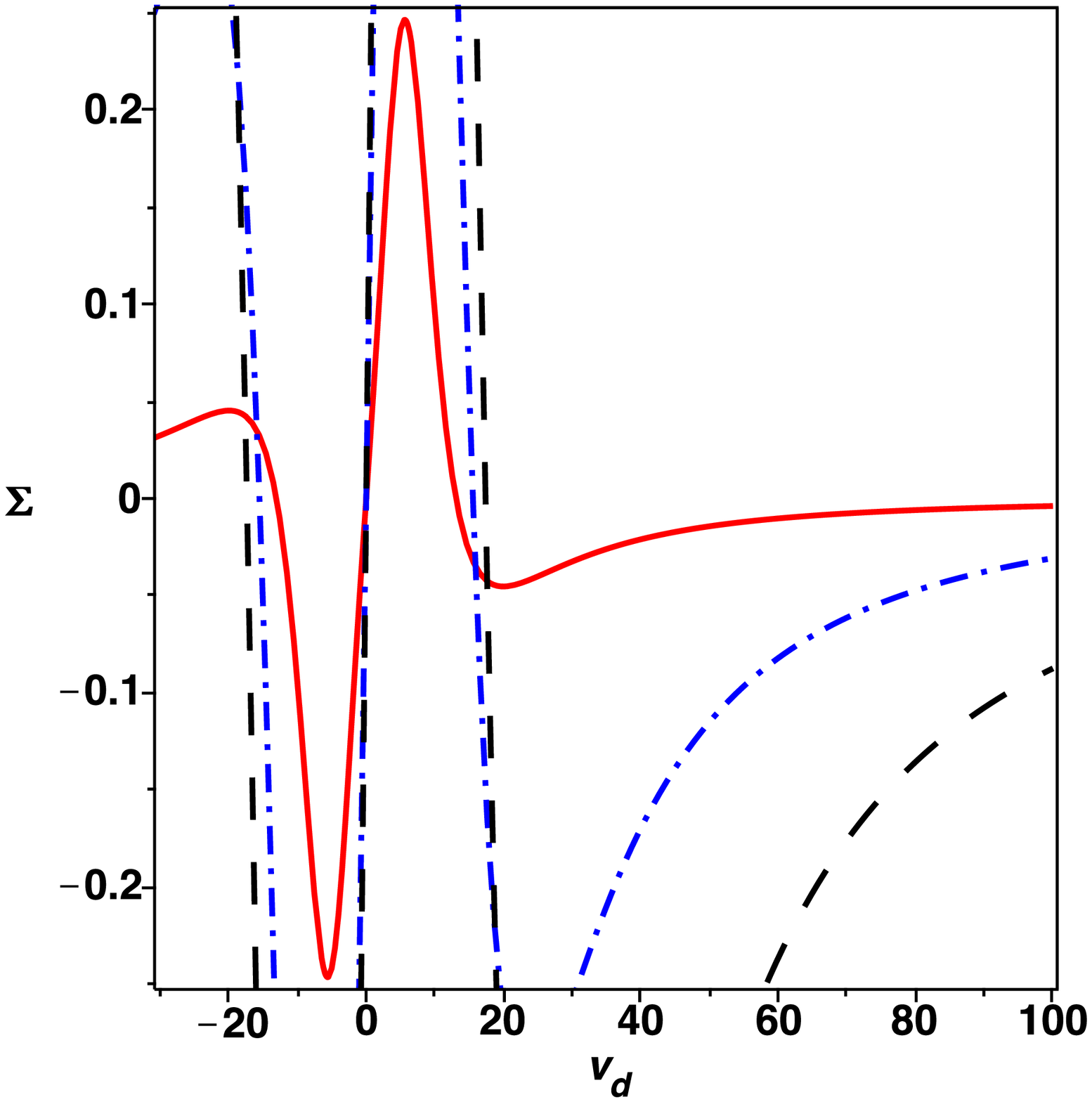}
  \label{fig:sfig1}
\end{subfigure}
\caption{On the left panel is the value of the integral in equation (\ref{maineq4}) as a function of the normalized drift velocity of the hot electron distribution function. The red, blue (dot-dashed), and black (dashed) lines are for temperature ratios $T_h/T_c=[10,50,100]$ respectively.  Negative (positive) integral values indicate parameter space for which the structure grows (damps) for coefficients $\alpha<0$ in equation (\ref{maineq4}). On the right panel is a zoom of the left panel showing the transition to negative value and spread for larger values of $T_h/T_c$.}  
\label{fig:fig}
\end{figure}

\noindent In Figure \ref{fig:fig} we plot the value of the integral in equation (\ref{maineq4}), without the argument $\alpha \sqrt{\Phi}/\bar{\tau}$, as a function of the normalized drift velocity of the hot electron distribution function. The red, blue (dot-dashed), and black (dashed) lines are for temperature ratios $T_h/T_c=[10,50,100]$ respectively. A zoom of the right panel shows the transition to negative values (i.e. growth of structures for $\alpha<0$ and damping for $\alpha>0$). We notice that for zero drift between the cold and hot component, the integral becomes zero, and the structure can neither grow, nor damp. As we shift the hot electron distribution, the integral alternates between positive and negative values, which translate respectively to damping (for ranges $0<v_d^h/v_{tc}< \ \sim 12$ and $v_d^h/v_{tc}>-13$) and growth (for $0>v_d^h/v_{tc}> \ \sim -12$ and $v_d^h/v_{tc}>13$) of $\delta f$ for $\alpha<0$. For $\alpha>0$, the sign of $\gamma$ is simply switched for the same intervals. The basic condition for temporal evolution of the structure relies on an asymmetric broadening of the hot electron distribution for $v_d\neq 0$. For zero drift between the cold and hot components, the even diffusion coefficient broadens the hot electron component symmetrically around zero with no net momentum transfer. We also notice from Figure \ref{fig:fig} that as we increase the temperature of the hot component with respect to the cold component, a greater number of hot electrons can be scattered, enhancing both growth and damping for a given drift velocity $v_d$.

\begin{figure}
\begin{subfigure}{.26\textwidth}
  \centering
  \includegraphics[height=4.2cm,width=3.5cm]{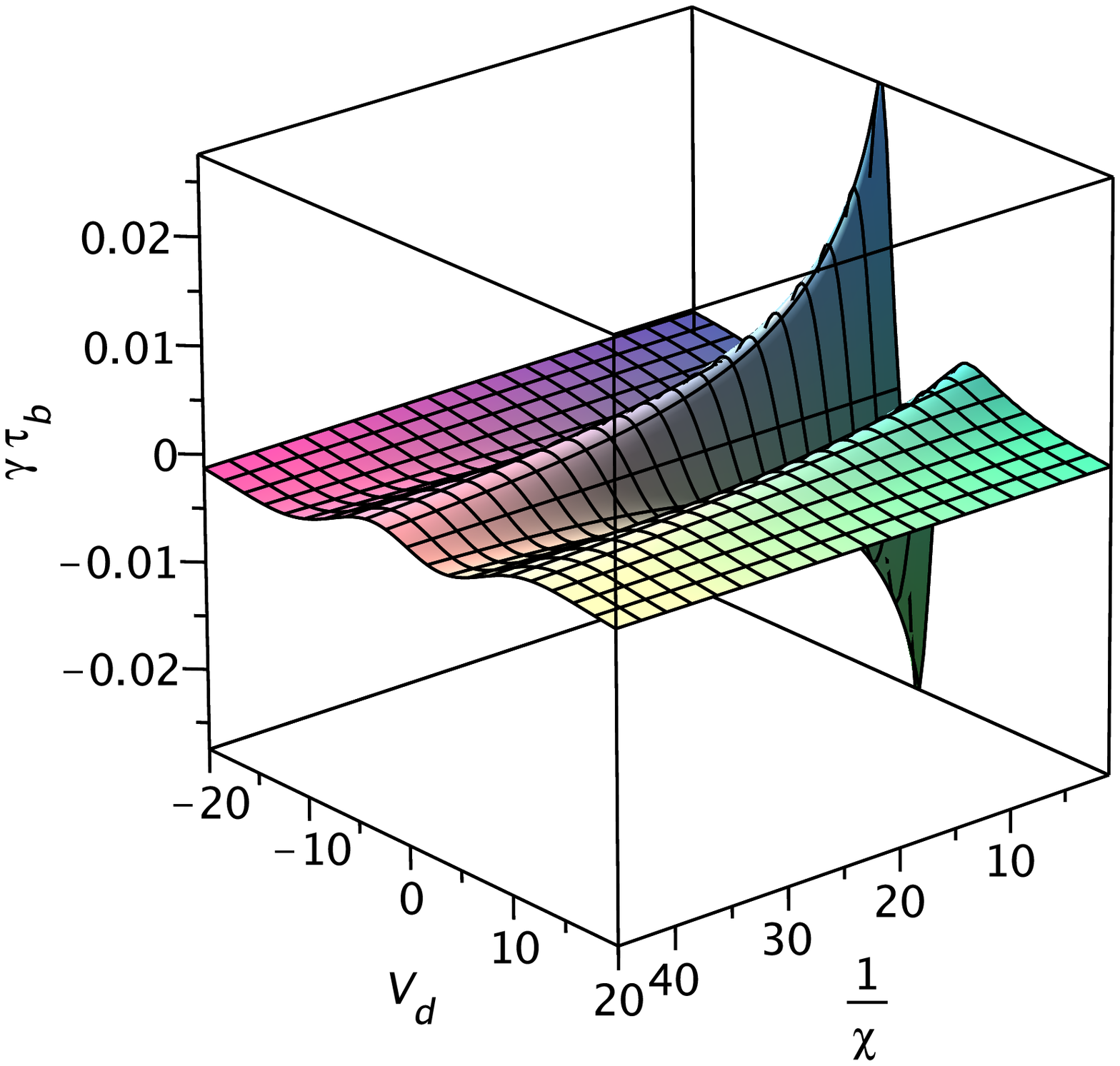}
  \label{fig:sfig1}
\end{subfigure}%
\begin{subfigure}{.26\textwidth}
  \centering
\includegraphics[height=4.2cm,width=3.5cm]{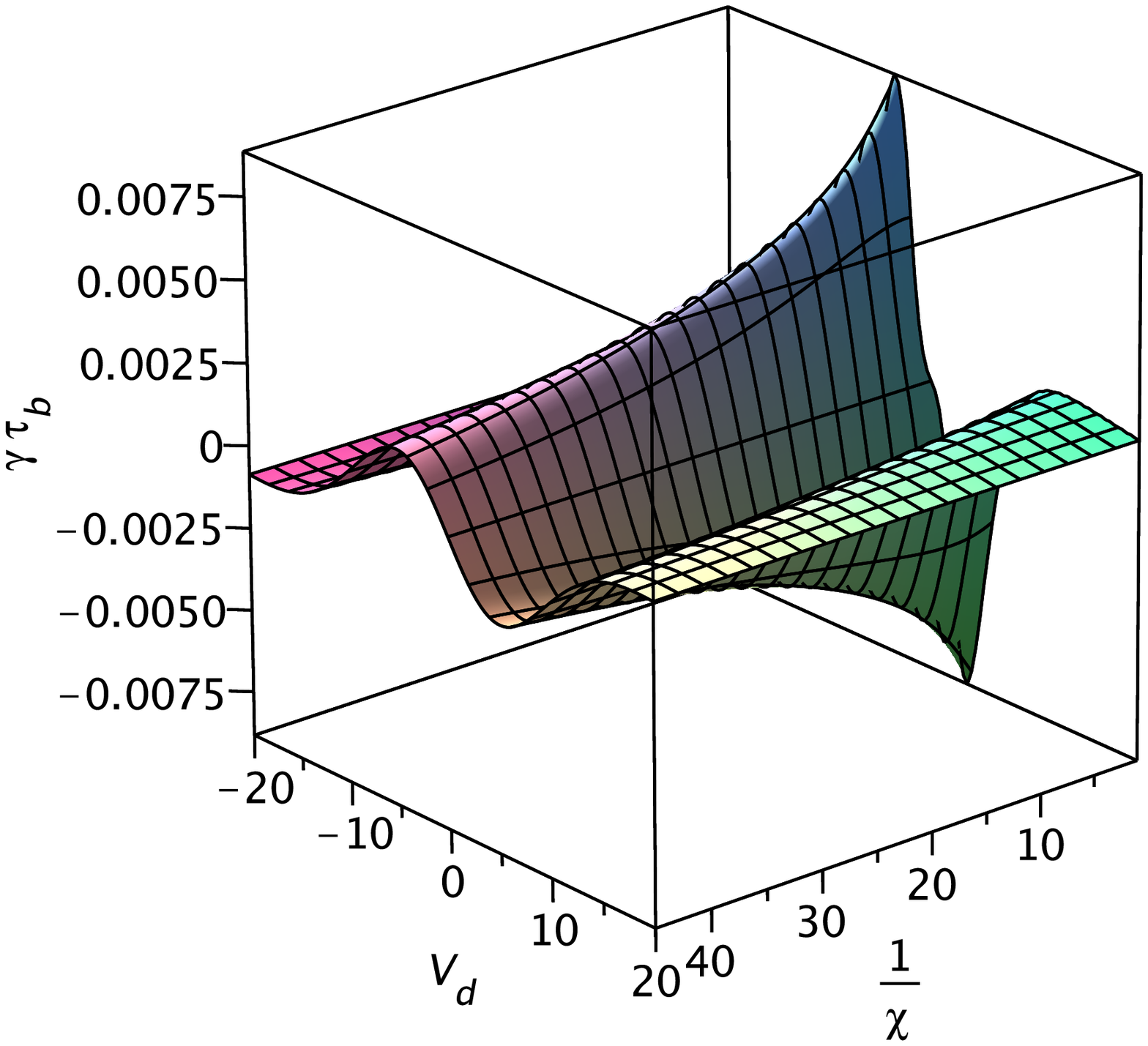}
  \label{fig:sfig2}
\end{subfigure}
\begin{subfigure}{.26\textwidth}
  \centering
  \includegraphics[height=4.2cm,width=3.5cm]{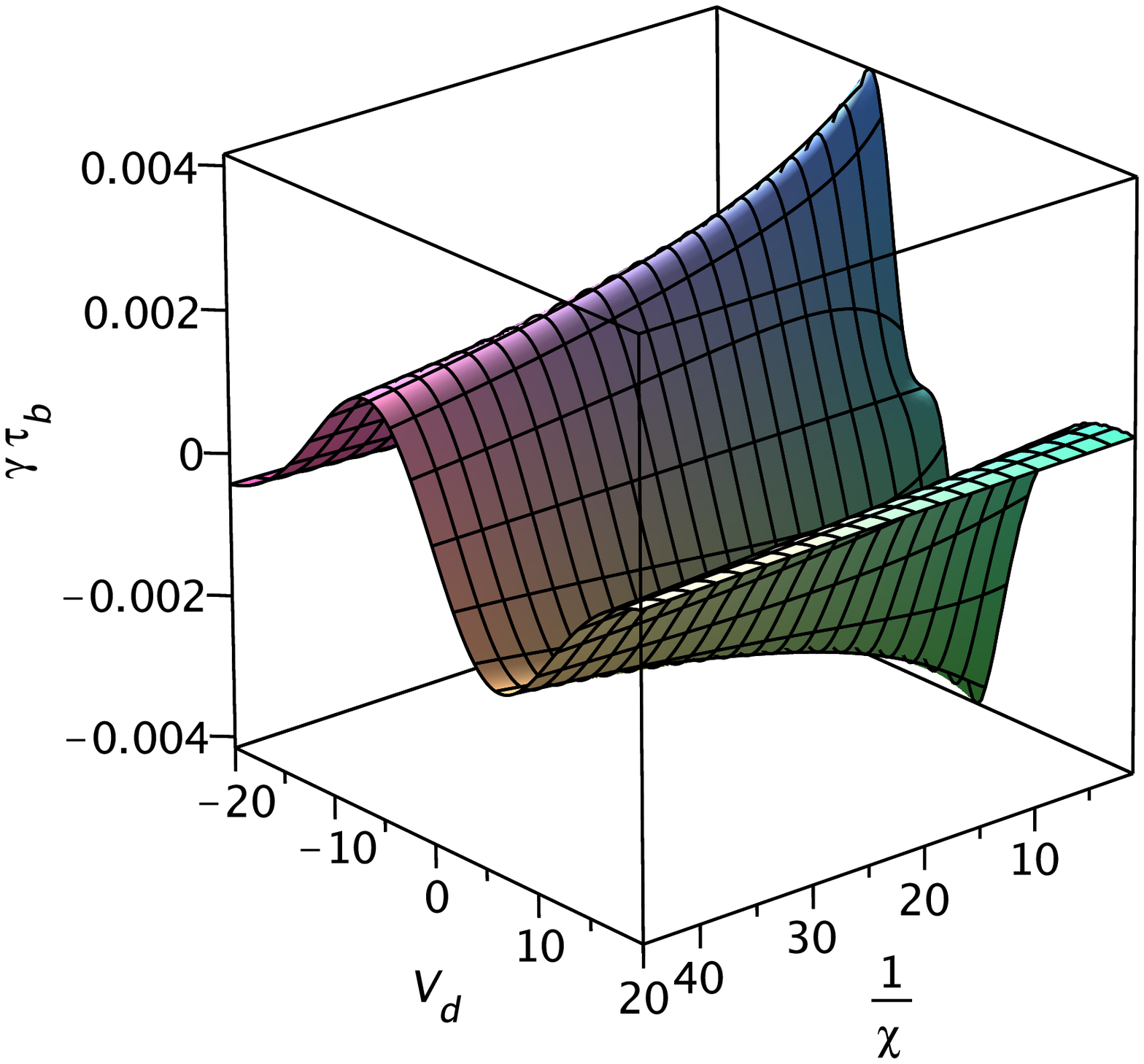}
 
  \label{fig:sfig3}
\end{subfigure}%
\begin{subfigure}{.26\textwidth}
  \centering
  \includegraphics[height=4.2cm,width=3.5cm]{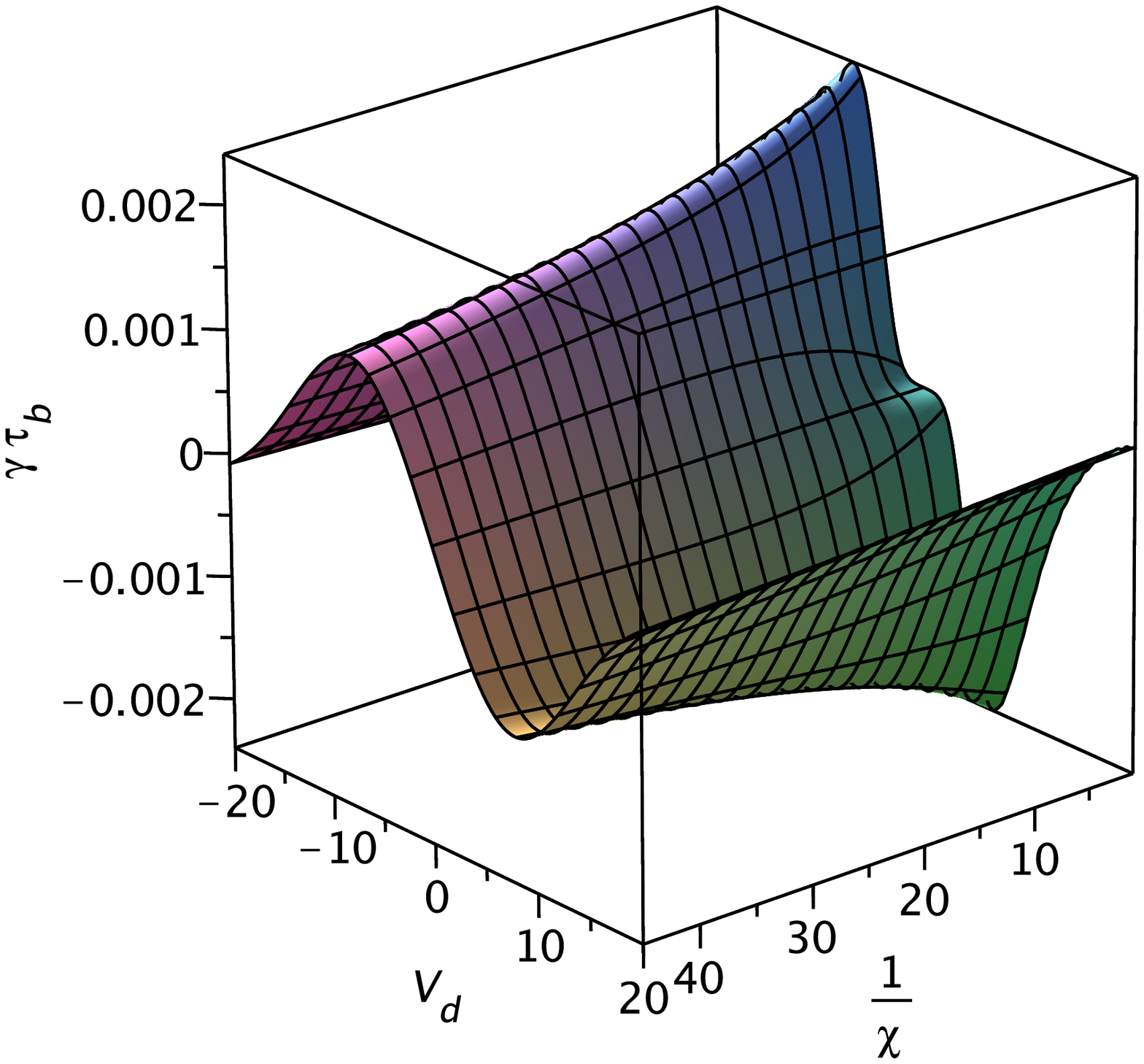}
  \label{fig:sfig4}
\end{subfigure}%
\caption{Dependence of the instantaneous growth rate $\gamma$ on the temperature ratio $\chi=T_c/T_h$ ranging between 1 and $40$, and normalized drift speed $v_d/v_{tc}$ ranging between 0 and 20. The density is fixed to $n_h/n_c=10^{-3}$, the normalized electrostatic potential to $\Phi=0.1$ and we assumed a large number of interactions with $\bar{\tau}=\tau_b/N$ with $N=1000$. The growth rate is normalized by the bounce frequency $\tau_b$. Figures from the left to the right were plotted for $d/\lambda_D=[2,4,6,8]$.}
\label{fig:fig2}
\end{figure}

\noindent Since the growth rate is dependent on the amplitude of the phase-space structure, it is not a constant like in linear theory. The electric potential $q\phi$ and the phase-space hole evolve as $\phi, \delta f \propto \exp(\sqrt{\phi}t)$ and saturate when the momentum associated with the structure in the frame of the cold electron goes to zero, i.e., $x=u/v_{tc} \simeq 0$, or when the drift between the hot and cold electrons is of the order of $v_d/v_{tc}<1$. Keeping these specificities of the subcritical instability in mind, it is useful to look at the scaling of the growth rate $\gamma$ for typical radiation belts conditions. In Figure (\ref{fig:fig2}) the growth rate is plotted as a function of the temperature ratio $\chi=T_c/T_h$ ranging between 1 and $40$, and normalized drift speed $v_d/v_{tc}$ ranging between -20 and 20. The density is fixed to $n_h/n_c=10^{-3}$, the normalized electrostatic potential to $\Phi=0.1$ and we assumed a large number of interactions with $\bar{\tau}=\tau_b/N$ with $N=1000$. The growth rate is normalized by the bounce frequency $\tau_b$. Figures from the left to the right were plotted for $d/\lambda_D=[2,4,6,8]$. The dependence of the growth rate on the average scale of the structure is not trivial since it enters the integral. Yet we note that the growth has similar scaling for average spatial scales ranging between $2$ and $8$ Debye length. Figure 2 highlights the fact that the described mechanism requires a large-number of interactions between the hot electrons and the phase-space structures to operate efficiently. Consequently, phase-space holes would need to be long-lived. Mechanisms in planetary radiation belts resulting in depletion (enhancement) of the hot electron population would slow down (enhance) phase-space structure growth.  Hence, we expect pitch-angle scattering of hot electrons outside of the radiation belts by whistlers to reduce the efficiency of the mechanism, and injection in the inner magnetosphere of hot electrons due to substorms to enhance growth rate by modification of the hot to cold electron density ratio.

\subsection{Comparisons with linear theory of electron acoustic mode}
\noindent It is useful to compare our results to the linear theory of electron-acoustic modes for similar plasma conditions, i.e., homogeneous, unmagnetized, one dimensional kinetic plasma with two electron populations described by Maxwellians. \cite{Gary87} determined that electron-acoustic modes could grow for hot electrons drifting with speeds $v_d > 5 v_{tc}$, temperature ratios $T_h/T_c \ge 10$ and hot and cold densities of comparable values, i.e. $n_h \simeq n_c$. While these conditions are consistent with our results, we find that structure growth for density ratios of $n_h/n_c \sim 10^{-3}$ remains a possibility. Contrary to the results of \cite{Gary87}, who find that structures with electron-acoustic properties are heavily damped for a small relative drift between the hot and cold components, a very small drift $v_d/v_{tc} \simeq 1$ for $T_c/T_h\simeq 10-50$ can also result in nonlinear growth. An equivalent discrepancy between linear and nonlinear instability has long been noted noted for the current-driven ion-acoustic mode \citep{Diamond15}. In the absence of linear instabilities, phase-space structures can mediate momentum-exchange between different plasma species in planetary magnetosphere environments. 

\section{Conclusion}
\noindent We have developed a heuristic model to quantify the nonlinear growth of phase-space electron holes in an electrostatic plasma with hot and cold electron components. Phase-space structures are driven by momentum exchange between hot and cold electron populations, and can grow for linearly stable plasmas. For a parameter space comparable with the Earth's radiation belts, with $T_h/T_c \simeq 10$, and $n_h/n_c \simeq 10^{-3}$, growth rates scale as $\gamma \bar{\tau_b} \simeq 10^{-2}N \sqrt{q\Phi/T_c}$, in which $N$ stands for the average number of interaction between hot electrons and phase-space structures over one bounce period $\tau_b$. The primary requirement for subcritical phase-space structures with electron-acoustic properties to grow is for the drift velocity $v_d \geq v_{tc}$ between the cold and hot electron component to be different from zero. This condition is much less stringent than for the linear electron acoustic mode which requires $v_d> 5 v_{tc}$ and should be commonly met for magnetically confined populations in planetary magnetospheres. For instance, various wave-particle interactions and energetic electrons injected after substorms in the inner magnetosphere can  translate into drifts between energetic and cold electrons on very short electron timescales. Observationally, the existence of hundreds of phase-space holes' electrostatic signature is a strong indication that electron currents are commonly generated in the inner magnetosphere. It is interesting to note that \textit{in situ} measurement in the inner magnetosphere of electron distribution functions with long-lived phase-space structures preceded the Van Allen Probes discovery \cite{Kellog10,Wilson11}. An appropriate description of kinetic electrons in the Earth's radiation belts might require incorporating time scales smaller than the bounce period where nonlinearities (trapping, coherent structure formation and growth, microbursts) have been shown to occur. Future studies will extend the current model to a magnetized plasma and take into account momentum exchange with whistler turbulence. Although the self-consistent problem describing the interactions of many holes with various electron populations will require a numerical approach, simpler, more tractable models, such as the one presented herein, have been proven useful to gain intuitive understanding of nonlinear phenomena in collisionless plasmas \citep{Berk90, Lesur13}.\\


\noindent Acknowledgements: A.O., A.P.D. and T.I.P. research was supported by Academy of Finland grants $\#267073/2013$ and $\#297688/2015$.

\bibliographystyle{jpp}


\end{document}